\documentclass[twocolumn,floatfix]{revtex4}
\usepackage{graphicx}
\usepackage{amsmath}
\usepackage{amssymb}
\usepackage{bm}

\begin{document}

\title{X-shaped and Y-shaped Andreev resonance profiles in a superconducting quantum dot}
\author{Shuo Mi}
\affiliation{Instituut-Lorentz, Universiteit Leiden, P.O. Box 9506, 2300 RA Leiden, The Netherlands}
\author{D. I. Pikulin}
\affiliation{Instituut-Lorentz, Universiteit Leiden, P.O. Box 9506, 2300 RA Leiden, The Netherlands}
\author{M. Marciani}
\affiliation{Instituut-Lorentz, Universiteit Leiden, P.O. Box 9506, 2300 RA Leiden, The Netherlands}
\author{C. W. J. Beenakker}
\affiliation{Instituut-Lorentz, Universiteit Leiden, P.O. Box 9506, 2300 RA Leiden, The Netherlands}
\date{May 2014}
\begin{abstract}
The quasi-bound states of a superconducting quantum dot that is weakly coupled to a normal metal appear as resonances in the Andreev reflection probability, measured via the differential conductance. We study the evolution of these Andreev resonances when an external parameter (such as magnetic field or gate voltage) is varied, using a random-matrix model for the $N\times N$ scattering matrix. We contrast the two ensembles with broken time-reversal symmetry, in the presence or absence of spin-rotation symmetry (class C or D). The poles of the scattering matrix in the complex plane, encoding the center and width of the resonance, are repelled from the imaginary axis in class C. In class D, in contrast, a number $\propto\sqrt{N}$ of the poles has zero real part. The corresponding Andreev resonances are pinned to the middle of the gap and produce a zero-bias conductance peak that does not split over a range of parameter values (Y-shaped profile), unlike the usual conductance peaks that merge and then immediately split (X-shaped profile).\\
{\tt Contribution for the JETP special issue in honor of A.F. Andreev's 75th birthday.}
\end{abstract}
\maketitle

\section{Introduction}
\label{intro}

\begin{figure}[tb]
\centerline{\includegraphics[width=0.8\linewidth]{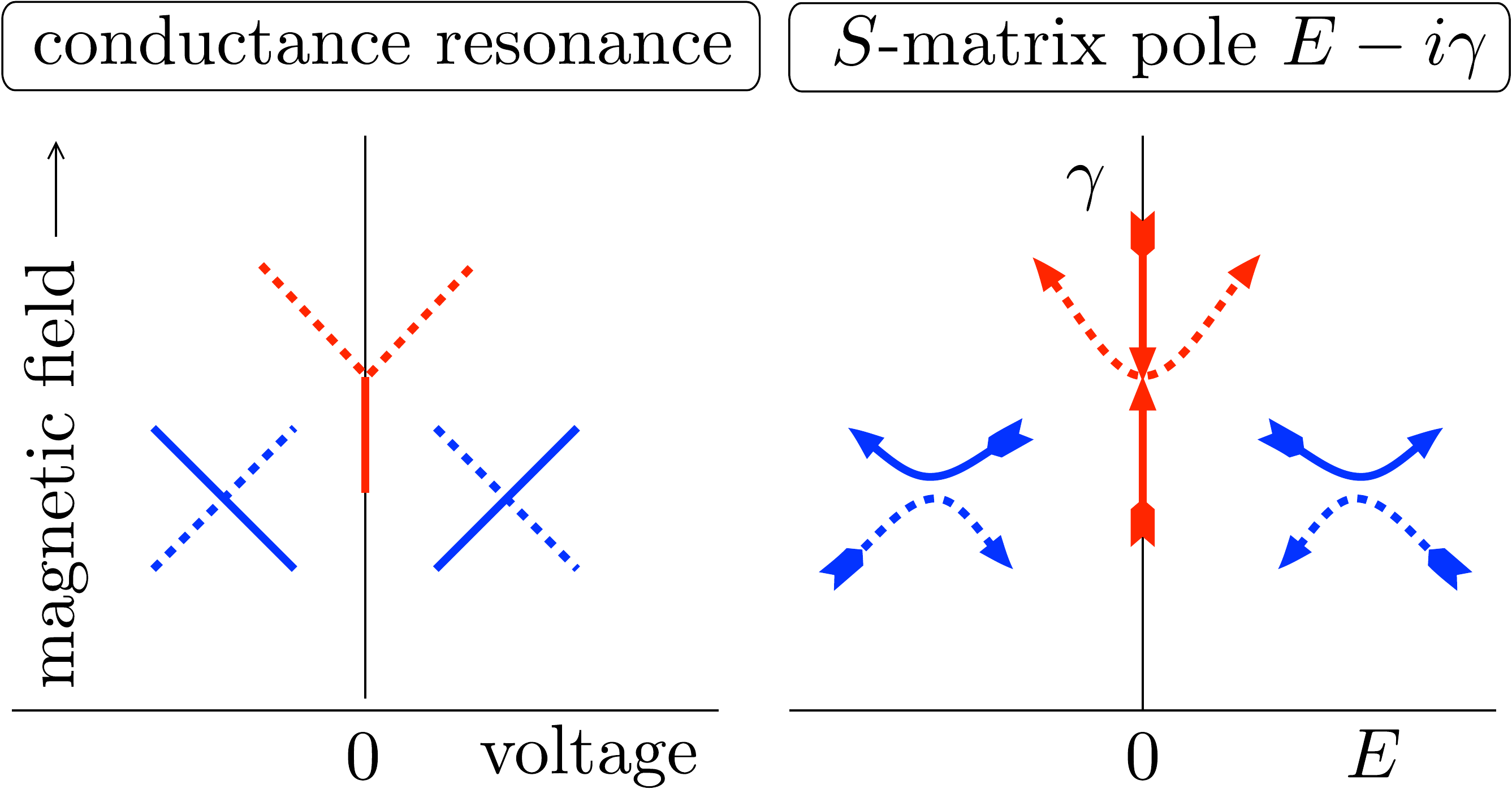}}
\caption{Left panel: Magnetic field $B$-dependence of peaks in the differential conductance $G=dI/dV$. The peak positions trace out an X-shaped or Y-shaped profile in the $B$-$V$ plane. Right panel: Location of the poles of the scattering matrix $S(\varepsilon)$ in the complex energy plane $\varepsilon=E-i\gamma$. The arrows indicate how the poles moves with increasing magnetic field.
}
\label{fig_XY}
\end{figure}

Half a century has passed since Alexander Andreev reported the curious retro-reflection of electrons at the interface between a normal metal and a superconductor \cite{And64}. One reason why Andreev reflection is still very much a topic of active research, is the recent interest in Majorana zero-modes \cite{Sil14}: Nondegenerate bound states at the Fermi level ($E=0$) consisting of a coherent superposition of electrons and holes, coupled via Andreev reflection. These are observed in the differential conductance as a resonant peak around zero bias voltage $V$ that does not split upon variation of a magnetic field $B$ \cite{Ali12,Lei12,Sta13,Bee13b}. In the $B,V$ plane the conductance peaks trace out an unusual Y-shaped profile, distinct from the more common X-shaped profile of peaks that meet and immediately split again. (See Fig.\ \ref{fig_XY}.)

It is tempting to think that the absence of a splitting of the zero-bias conductance peak demonstrates that the quasi-bound state is nondegenerate, hence Majorana. This is mistaken. As shown in a computer simulation \cite{Pik12}, the Y-shaped conductance profile is generic for superconductors with broken spin-rotation and broken time-reversal symmetry, irrespective of the presence or absence of Majorana zero-modes. The theoretical analysis of Ref.\ \onlinecite{Pik12} focused on the ensemble-averaged conductance peak, in the context of the weak antilocalization effect \cite{Bro95,Alt96,Ios12,Bag12}. Here we analyse the sample-specific conductance profile, by relating the X-shape and Y-shape to different configurations of poles of the scattering matrix in the complex energy plane \cite{Pik11}. 

\section{Andreev billiard}
\label{Andreevbilliard}

\begin{figure}[tb]
\centerline{\includegraphics[width=0.7\columnwidth]{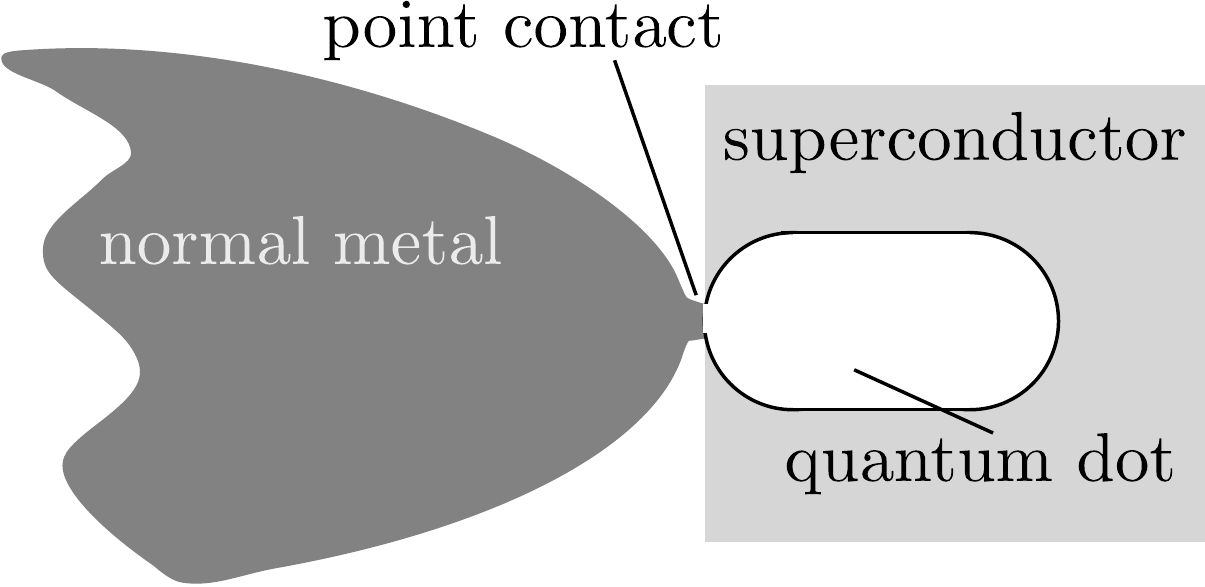}}
\caption{Schematic illustration of an Andreev billiard.}
\label{fig:SystemSetup}
\end{figure}

\subsection{Scattering resonances}
\label{scatres}

We study the Andreev billiard geometry of Fig.\ \ref{fig:SystemSetup}: A semiconductor quantum dot strongly coupled to a superconductor and weakly coupled to a normal metal. In the presence of time-reversal symmetry an excitation gap is induced in the quantum dot by the proximity effect \cite{Bee04}. We assume that the gap is closed by a sufficiently strong magnetic field. Quasi-bound states can then appear near the Fermi level ($E=0$), described by the Hamiltonian
\begin{equation}
{\cal H} =  \sum_{\mu, \nu} |\mu\rangle H_{\mu \nu} \langle\nu| + \sum_{\mu, a} \bigl( |\mu\rangle W_{\mu a} \langle a| + |a\rangle W^{\ast}_{\mu a} \langle\mu| \bigr). \label{Eq01}
\end{equation}
The bound states in the closed quantum dot are eigenvalues of the $M\times M$ Hermitian matrix $H=H^{\dagger}$. The $M\times N$ matrix $W$ couples the basis states $|\mu\rangle$ in the quantum dot to the normal metal, via $N$ propagating modes $|a\rangle$ through a point contact. In principle we should take the limit $M\rightarrow\infty$, but in practice $M\gg N$ suffices.

The amplitudes of incoming and outgoing modes in the point contact at energy $E$ (relative to the Fermi level) are related by the $N\times N$ scattering matrix \cite{Guh98,Bee97}
\begin{equation}
S(E) = 1 + 2 \pi i W^{\dagger} \left(H - i \pi WW^{\dagger}-E\right)^{-1}W.\label{Eq03}
\end{equation}
This is a unitary matrix, $S(E)S^{\dagger}(E)=1$.

A scattering resonance corresponds to a pole $\varepsilon=E-i\gamma$ of the scattering matrix in the complex energy plane, which is an eigenvalue of the non-Hermitian matrix 
\begin{equation}
H_{\rm eff}=H - i \pi WW^{\dagger}.\label{Heffdef}
\end{equation}
The positive definiteness of $WW^{\dagger}$ ensures that the poles all lie in the lower half of the complex plane, $\gamma\geq 0$, as required by  causality. Particle-hole symmetry implies that $\varepsilon$ and $-\varepsilon^{\ast}$ are both eigenvalues of $H_{\rm eff}$, so the poles are symmetrically arranged around the imaginary axis.

The differential conductance $G(V)=dI/dV$ of the quantum dot, measured by grounding the superconductor and applying a bias voltage to the normal metal, is obtained from the scattering matrix via \cite{Pik12}
\begin{align}
G(V)=\frac{e^{2}}{h}\left[\frac{N}{2}-\frac{1}{2}\,{\rm Tr}\,S(eV)\tau_z S^\dagger(eV)\tau_z\right],\label{Gehbasis}
\end{align}
in the electron-hole basis, and
\begin{align}
G(V)=\frac{e^{2}}{h}\left[\frac{N}{2}-\frac{1}{2}\,{\rm Tr}\,S(eV)\tau_y S^\dagger(eV)\tau_y\right],\label{GMbasis}
\end{align}
in the Majorana basis. The Pauli matrices $\tau_y$, $\tau_z$ act on the electron-hole degree of freedom. The two bases are related by the unitary transformation
\begin{equation}
S\mapsto USU^{\dagger},\;\;U=\sqrt{\tfrac{1}{2}}\begin{pmatrix}
1&1\\
i&-i
\end{pmatrix}.\label{Udef}
\end{equation}

\subsection{Gaussian ensembles}
\label{gaussens}

For a random-matrix description we assume that the scattering in the quantum dot is chaotic, and that this applies to normal scattering from the electrostatic potential as well as to Andreev scattering from the pair potential. In the large-$M$ limit we may then take a Gaussian distribution for $H$,
\begin{equation}
P(H)\propto \exp\left(-\frac{c}{M} \,{\rm Tr}\,H^2\right).\label{Eq02}
\end{equation}

By taking the matrix elements of $H$ to be real, complex, or quaternion numbers (in an appropriate basis), one obtains the Wigner-Dyson ensembles of non-superconducting chaotic billiards \cite{Mehta,Forrester,handbook}. Particle-hole symmetry then plays no role, because normal scattering does not couple electrons and holes. 

Altland and Zirnbauer introduced the particle-hole symmetric ensembles appropriate for an Andreev billiard \cite{Alt97}. The two ensembles without time-reversal symmetry are obtained by taking the matrix elements of $i\times H$ (instead of $H$ itself) to be real or quaternion. When $iH$ is real there is only particle-hole symmetry (class D), while when $iH$ is quaternion there is particle-hole and spin-rotation symmetry (class C). 

Both the Wigner-Dyson (WD) and the Altland-Zirnbauer (AZ) ensembles are characterized by a parameter $\beta\in\{1,2,4\}$ that describes the strength of the level repulsion factor  in the probability distribution of distinct eigenvalues $E_i$ of $H$: a factor $\prod_{i<j}|E_i-E_j|^\beta$ in the WD ensembles and a factor $\prod'_{i<j}|E_i^2-E_j^2|^\beta$ in the AZ ensembles. (The prime indicates that the product includes only the positive eigenvalues.)

In the WD ensembles the parameter $\beta$ also counts the number of degrees of freedom of the matrix elements of $H$: $\beta=1$, 2 or 4 when $H$ is real, complex, or quaternion, respectively. In the AZ ensembles this connection is lost: $\beta=2$ in the class C ensemble ($iH$ real) as well as in the class D ensemble ($iH$ quaternion).

The coefficient $c$ can be related to the average spacing $\delta_0$ of distinct eigenvalues of $H$ in the bulk of the spectrum, 
\begin{equation}
c=\frac{\beta\pi^{2}}{8\delta_0^{2}}\times\begin{cases}
2& \text{in the WD ensembles},\\
1& \text{in the AZ ensembles.}
\end{cases}\label{cdef}
\end{equation}
The coefficient \eqref{cdef} for the AZ ensembles is twice as small as it is in the WD ensembles with the same $\beta$, on account of the $\pm E$ symmetry of the spectrum, see App.\ \ref{cpmEsym}.

Because the distribution of $H$ is basis independent, we may without loss of generality choose a basis such that the coupling matrix $W$ is diagonal,
\begin{equation}
W_{mn}=w_n \delta_{mn},\;\;1\leq m\leq M,\;\;1\leq n\leq N.\label{Wdef}
\end{equation}
The coupling strength $w_n$ is related to the tunnel probability $\Gamma_n\in (0,1)$ of mode $n$ into the quantum dot by \cite{Guh98,Bee97}
\begin{equation}
|w_n|^2 = \frac{M\delta_0}{\pi^2 \Gamma_n}\bigl( 2 - \Gamma_n - 2 \sqrt{1-\Gamma_n} \bigr).\label{Eq04}
\end{equation}

\subsection{Class C and D ensembles}
\label{symclassCD}

We summarize the properties of the $\beta=2$ Altland-Zirnbauer ensembles, symmetry class C and D, that we will need for our study of the Andreev resonances. (See App.\ \ref{CIDIIIensembles} for the corresponding $\beta=1,4$ formulas in symmetry class CI and DIII.) Similar formulas can be found in Ref.\ \onlinecite{Iva02}.

When Andreev scattering operates together with spin-orbit coupling, one can combine electron and hole degrees of freedom from the same spin band into a real basis of Majorana fermions. [This change of basis amounts to the unitary transformation \eqref{Udef}.] In the Majorana basis the constraint of particle-hole symmetry reads simply
\begin{equation}
H=-H^{\ast},\label{Dconstraint}
\end{equation}
so we can take $H=iA$ with $A$ a real antisymmetric matrix. In the Gaussian ensemble the upper-diagonal matrix elements $A_{nm}$ ($n<m$) all have identical and independent distributions,
\begin{equation}
P(\{A_{nm}\})\propto\prod_{1=n<m}^{M}\exp\left(-\frac{\pi^{2}A_{nm}^{2}}{2M\delta_0^{2}}\right),\label{GaussEns}
\end{equation}
see Eqs.\ \eqref{Eq02} and \eqref{cdef}. This is the $\beta=2$ class-D ensemble, without spin-rotation symmetry.

The $\beta=2$ class-C ensemble applies in the absence of spin-orbit coupling, when spin-rotation symmetry is preserved. Andreev reflection from a spin-singlet superconductor couples only electrons and holes from different spin bands, which cannot be combined into a real basis state. It is then more convenient stay in the electron-hole basis and to eliminate the spin degree of freedom by considering a single spin band for the electron and the opposite spin band for the hole. (The matrix dimensionality $M$ and the mean level spacing $\delta_0$ then refer to a single spin.) In this basis the particle-hole symmetry requires
\begin{equation}
H=-\tau_y H^\ast \tau_y,\label{Cconstraint} 
\end{equation}
where the Pauli matrix $\tau_y$ operates on the electron and hole degrees of freedom. 

The constraint \eqref{Cconstraint} implies that $H=iQ$ with $Q$ a quaternion anti-Hermitian matrix. Its matrix elements are of the form
\begin{equation}
\begin{split}
&Q_{nm}=a_{nm}\tau_0+ib_{nm}\tau_x+ic_{nm}\tau_y+id_{nm}\tau_z,\\
&n,m=1,2,\ldots M/2,
\end{split}
\label{quaterniondef}
\end{equation}
with real coefficients $a,b,c,d$ (to ensure that $Q_{nm}=\tau_y Q_{nm}^{\ast}\tau_y$). Anti-Hermiticity of $Q$ requires that the off-diagonal elements are related by $a_{nm}=-a_{mn}$ and $x_{nm}=x_{mn}$ for $x\in\{b,c,d\}$. On the diagonal $a_{nn}=0$. In the Gaussian ensemble the independent matrix elements have the distribution
\begin{align}
&P(\{Q_{nm}\})\propto\prod_{n=1}^{M/2}\exp\left(-\frac{\pi^{2}}{2M\delta_0^{2}}(b_{nn}^2+c_{nn}^2+d_{nn}^2)\right) \nonumber\\
&\quad\times  \prod_{1=n<m}^{M/2}\exp\left(-\frac{\pi^{2}}{M\delta_0^{2}}(a_{nm}^2+b_{nm}^2+c_{nm}^2+d_{nm}^2)\right),\label{GaussEnsC}
\end{align}

\section{Andreev resonances}
\label{Aresonances}

\subsection{Accumulation on the imaginary axis}
\label{accumulation}

\begin{figure*}[tb]
\centerline{\includegraphics[width=0.8\linewidth]{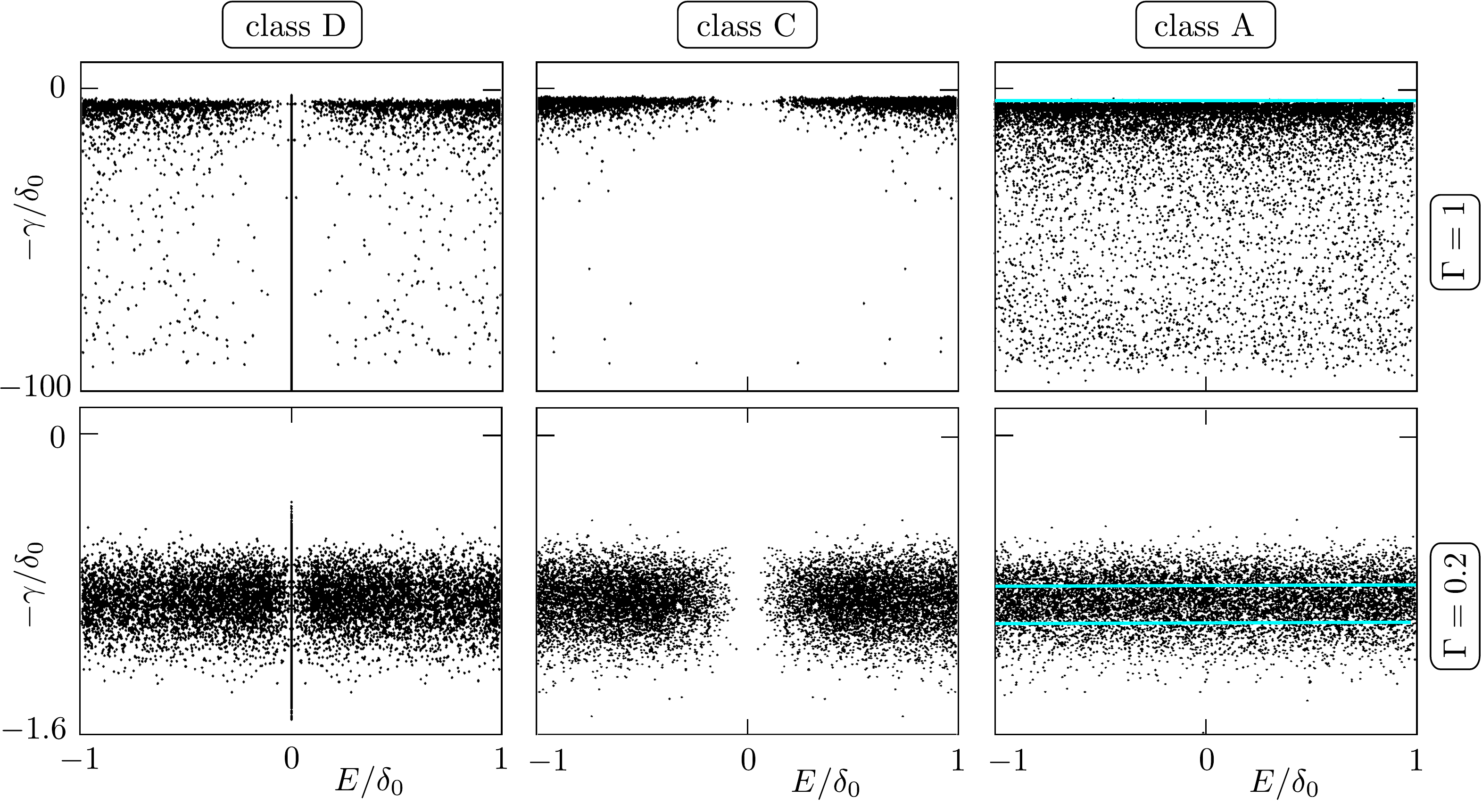}}
\caption{Scatter plot of the poles $\varepsilon=E-i\Gamma$ of 5000 scattering matrices $S(\varepsilon)$, in the Gaussian ensembles of class D, C, and A (first, second, and third column), for ballistic coupling ($\Gamma=1$, first row) and for tunnel coupling ($\Gamma=0.2$, second row). In each case the Hamiltonian has dimension $M\times M=500\times 500$ and the scattering matrix $N\times N=50\times 50$. Only a narrow energy range near $E=0$ is shown, to contrast the accumulation of the poles on the imaginary axis in class D and the repulsion in class C. The blue horizontal lines indicate the expected boundaries \eqref{rhoEgammab} of the class-A scatter plot in the limit $N,M/N\rightarrow\infty$.}
\label{fig_poles}
\end{figure*}

In Fig.\ \ref{fig_poles} we show the location of the poles of the scattering matrix in the complex energy plane, for the $\beta=2$ Altland-Zirnbauer ensembles with and without spin-rotation symmetry (class C and D, respectively). The $\beta=2$ Wigner-Dyson ensemble (class A, complex $H$) is included for comparison. The poles are eigenvalues $\varepsilon$ of the non-Hermitian effective Hamiltonian \eqref{Eq01}, with $H$ distributed according to the Gaussian distribution \eqref{Eq02}--\eqref{cdef}, $\beta=2$, and coupling matrix $W$ given by Eqs.\ \eqref{Wdef}--\eqref{Eq04}. For simplicity we took identical tunnel probabilities $\Gamma_n\equiv\Gamma$ for each of the $N$ modes connecting the quantum dot to the normal metal. 

The number $M$ of basis states in the quantum dot is taken much larger than $N$, to reach the random-matrix regime. In class C this number is necessarily even, as demanded by the particle-hole symmetry relation \eqref{Cconstraint}. The symmetry relation \eqref{Dconstraint} in class D imposes no such constraint, and when $M$ is odd there is an unpaired Majorana zero-mode in the spectrum \cite{Iva02,note3}. The class-D superconductor with a Majorana zero-mode is called topologically nontrivial, while class C or class D without a zero-mode is called topologically trivial \cite{Ryu10,Has10,Qi11}. For a more direct comparison of class C and class D we take $M$ even in both cases, so both superconductors are topologically trivial.

In the absence of particle-hole symmetry (class A), the poles $\varepsilon=E-i\gamma$ of the scattering matrix have a density \cite{Fyo97}
\begin{align}
&\rho(E,\gamma)=\frac{N}{4\pi\gamma^2},\;\;\gamma_{\rm min}<\gamma<\gamma_{\rm max},\label{rhoEgammaa}\\
&\gamma_{\rm min}=N\Gamma\delta_0/4\pi,\;\;\gamma_{\rm max}=\gamma_{\rm min}/(1-\Gamma),
\label{rhoEgammab}
\end{align}
for $|E|\ll M\delta_0$ and asymptotically in the limit $N,M/N\rightarrow\infty$. For $|E|\gtrsim\delta_0$ all three $\beta=2$ ensembles A, C, D have a similar density of poles, but for smaller $|E|$ the densities are strikingly different, see Fig.\ \ref{fig_poles}.  While in class C the poles are repelled from the imaginary axis, in class D they accumulate on that axis. 

As pointed out in Ref.\ \onlinecite{Pik11}, a nondegenerate pole $\varepsilon=-i\gamma$ on the imaginary axis has a certain stability, it cannot acquire a nonzero real part $E$ without breaking the $\varepsilon\leftrightarrow-\varepsilon^{\ast}$ symmetry imposed by particle-hole conjugation. To see why this stability is not operative in class C, we note that on the imaginary axis $\gamma$ is a real eigenvalue of the matrix
\begin{align}
&\Omega=-Q+\pi WW^{\dagger}\;\;\text{in class C},\label{OmegadefC}\\
&\Omega=-A+\pi WW^{\dagger}\;\;\text{in class D}.\label{OmegadefD}
\end{align}
In both classes the matrix $\Omega$ commutes with an anti-unitary operator, ${\cal C}\Omega=\Omega{\cal C}$, with ${\cal C}=i\tau_{y}{\cal K}$ in class C and ${\cal C}={\cal K}$ in class D. (The operator ${\cal K}$ performs a complex conjugation.) In class C this operator ${\cal C}$ squares to $-1$, so a real eigenvalue $\gamma$ of $\Omega$ has a Kramers degeneracy \cite{note2} and hence nondegenerate poles $\varepsilon=-i\gamma$ on the imaginary axis are forbidden. In class D, in contrast, the operator ${\cal C}$ squares to $+1$, Kramers degeneracy is inoperative and nondegenerate poles are allowed and in fact generic.

\subsection{Square-root law}
\label{squareroot}

\begin{figure}[tb]
\centerline{\includegraphics[width=0.8\linewidth]{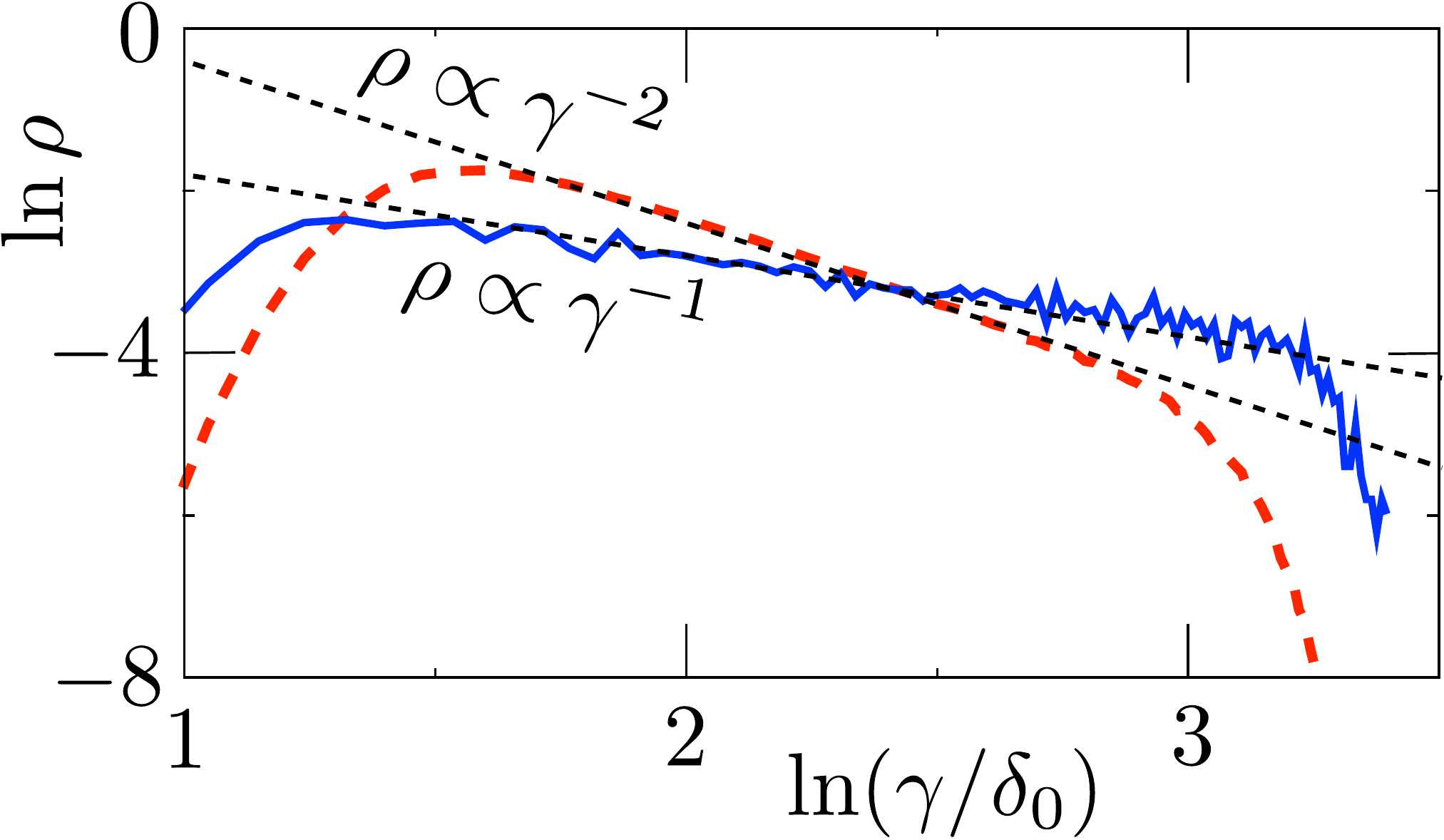}}
\caption{Double-logarithmic plot of the probability distribution $\rho(\gamma)$, normalized to unity, of the imaginary part $\gamma$ of the poles of the scattering matrix. The curves are calculated by averaging over some 2000 realizations of the class-D ensemble, with $N=10$, $M=500$, $\Gamma=0.9$. The red dashed curve includes all poles, while the blue solid curve includes only the poles on the imaginary axis ($E=0$). The black dashed lines are the predicted slopes from Eq.\ \eqref{rhoEgammaa} and \eqref{rho0result}.
}
\label{fig_densities}
\end{figure}

As we explain in App.\ \ref{orthogonalmapping}, for ballistic coupling ($\Gamma=1$) the statistics of poles on the imaginary axis can be mapped onto the statistics of the real eigenvalues of an $M\times M$ random orthogonal matrix with $N$ rows and columns deleted --- which is a solved problem \cite{Kho10,For10}. The linear density profile $\rho_0(\gamma)$ on the imaginary axis is
\begin{equation}
\rho_0(\gamma)=\sqrt{\frac{N\Gamma}{8\pi}}\,\frac{1}{\gamma},\;\;\gamma_{\rm min}<\gamma<\gamma_{\rm max},\label{rho0result}
\end{equation}
for $1\ll N\Gamma\ll M$ and $\gamma_{\rm min}$, $\gamma_{\rm max}$ given by Eq.\ \eqref{rhoEgammab}. We {\em conjecture} that this density profile, derived \cite{Kho10} for $\Gamma=1$, holds also for $\Gamma<1$. In Fig.\ \ref{fig_densities} we give numerical evidence in support of this conjecture.

\begin{figure}[tb]
\centerline{\includegraphics[width=0.8\linewidth]{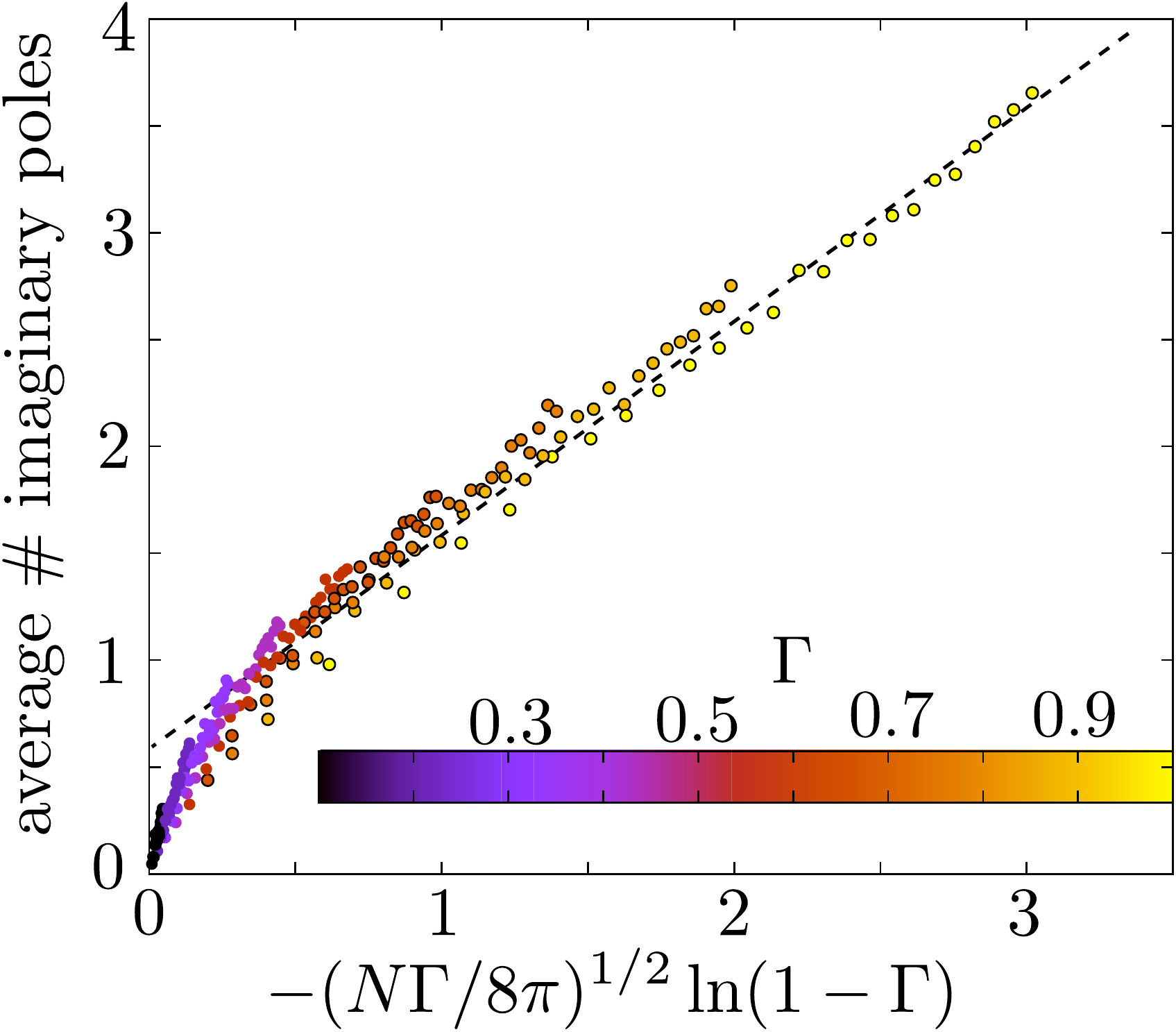}}
\caption{Average of the number $N_{\rm Y}$ of poles on the imaginary axis for an $N\times N$ scattering matrix $S(\varepsilon)$ in symmetry class D. Colors distinguish different tunnel couplings $\Gamma<1$, and $N$ is increased together with $M=80\,N$. The slope of the dashed black line is the large-$N$ asymptote \eqref{squarerootlaw}.}
\label{fig_polesgammacount}
\end{figure}

In Fig.\ \ref{fig_polesgammacount} we show how the average number $\langle N_{\rm Y}\rangle$ of class-D poles on the imaginary axis depends on the dimensionality $N$ of the scattering matrix and on the tunnel probability $\Gamma$. We compare with the square-root law \cite{note4}
\begin{equation}
\langle N_{\rm Y}\rangle=-\sqrt{\frac{N\Gamma}{8\pi}}\ln(1-\Gamma),\label{squarerootlaw}
\end{equation}
implied by integration of our conjectured density profile \eqref{rho0result}. This $\sqrt{N}$ scaling is generic for random-matrix ensembles that exhibit accumulation of eigenvalues on the real or imaginary axis, such as the Ginibre ensemble \cite{Gin65,Leh91,Ede94} (real Gaussian matrices without any symmetry) and the Hamilton ensemble \cite{Bee13} (matrices of the form ${\cal M}=HJ$ with $H$ a symmetric real Gaussian matrix and $J={\begin{pmatrix}
0&1\\
-1&0
\end{pmatrix}}$ a fixed anti-symmetric matrix). Fig.\ \ref{fig_polesgammacount} shows that the Andreev resonances follow the same square-root law.

\section{X-shaped and Y-shaped conductance profiles}

\begin{figure}[tb]
\centerline{\includegraphics[width=0.75\columnwidth]{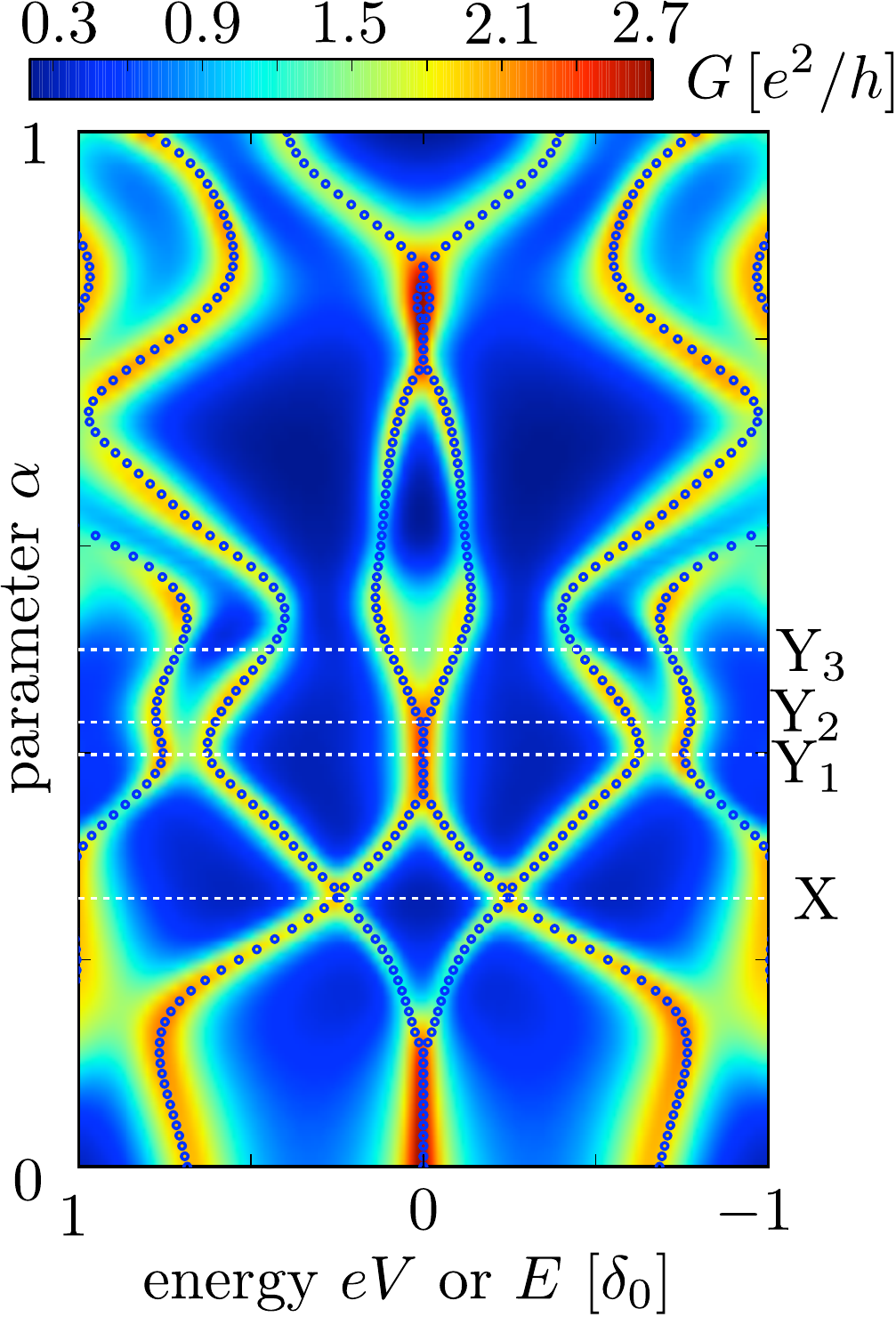}}
\caption{Parametric evolution of the differential conductance $G(V,\alpha)$ (color scale) and the real part $E$ of the poles of the scattering matrix $S_{\alpha}(\varepsilon)$. These are results for a single realization of the class D ensemble with $M=120$, $N=6$, and $\Gamma=0.3$.
} 
\label{fig:Pole_dynamics}
\end{figure}

\begin{figure}[tb]
\centerline{\includegraphics[width=0.8\columnwidth]{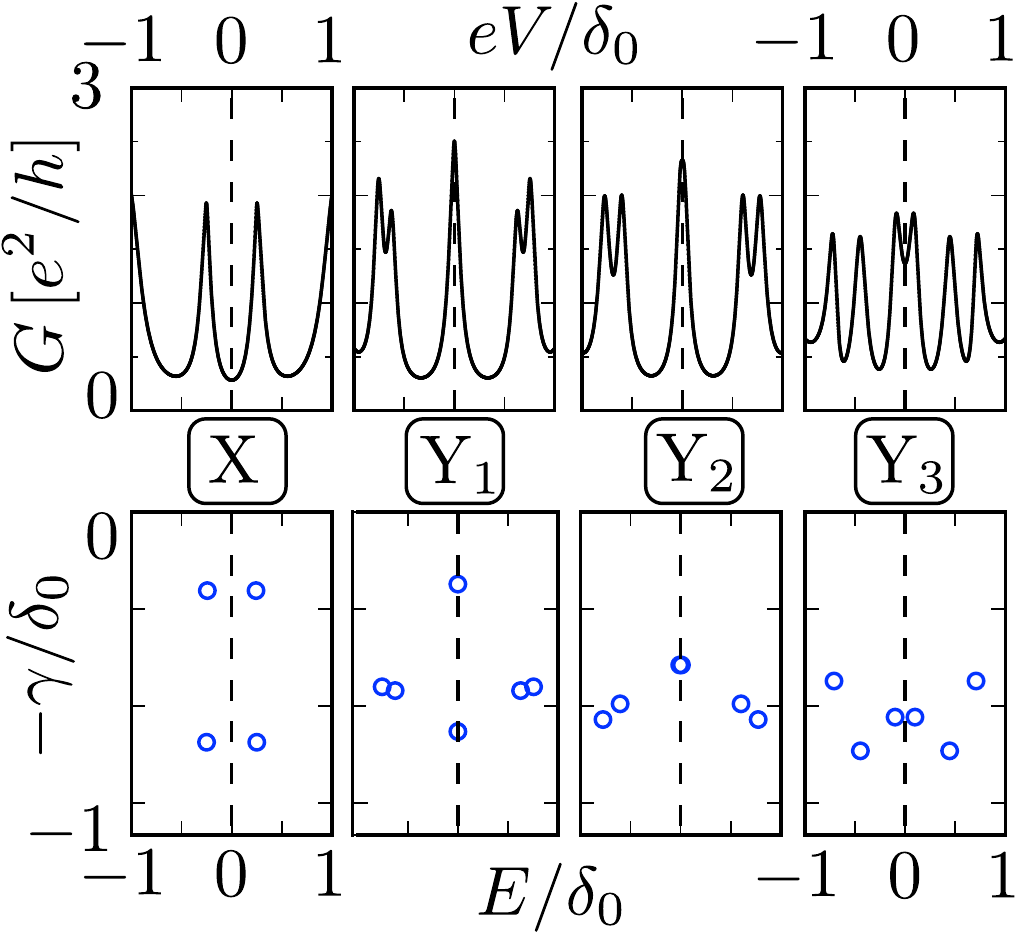}}
\caption{Four cuts through the parametric evolution of Fig.\ \ref{fig:Pole_dynamics}, showing the differential conductance $G=dI/dV$ (top row) and scattering matrix poles $\varepsilon=E-i\gamma$ (bottom row).
}
\label{fig:Cuts}
\end{figure}

In Ref.\ \onlinecite{Pik12} it was found in a computer simulation of a superconducting InSb nanowire that the conductance resonances trace out two distinct profiles in the voltage-magnetic field plane: an X-shape or a Y-shape. In the X-shaped profile a pair of conductance resonances merges and immediately splits again upon variation of voltage $V$ or magnetic field $B$. In the Y-shaped profile a pair of peaks merges at $V=0$ and then stays pinned to zero voltage over a range of magnetic field values. Here we wish to relate this phenomenology to the parametric evolution of poles of the scattering matrix in the complex energy plane \cite{Pik11}.

For that purpose we introduce a parameter dependence in the Hamiltonian $H$ of the Andreev billiard,
\begin{equation}
H_\alpha = (1-\alpha) H_0 + \alpha H_1, \label{Eq05}
\end{equation}
and calculate the differential conductance as a function of $V$ and $\alpha$. We work in symmetry class D (broken time-reversal and broken spin-rotation symmetry), so $H_0$ and $H_1$ are purely imaginary antisymmetric matrices (in the Majorana basis). We draw them from the Gaussian distribution \eqref{GaussEns}. The scattering matrix $S_\alpha$, obtained from $H_\alpha$ via Eq.\ \eqref{Eq03}, gives the differential conductance $G(V,\alpha)$ via Eq.\ \eqref{GMbasis}. For each $\alpha$ we also compute the poles $\varepsilon=E-i\gamma$ of $S(\varepsilon)$ in the complex energy plane.

Fig.\ \ref{fig:Pole_dynamics} shows a typical realization where the number $N_{\rm Y}$ of conductance poles on the imaginary axis switches between zero and two when $\alpha$ varies in the interval $[0,1]$. The color-scale plot shows $G(V,\alpha)$, while the dots trace out the projection of the poles of $S_{\alpha}(\varepsilon)$ on the real axis. Labels X and Y indicate the two types of profiles, and Fig.\ \ref{fig:Cuts} shows the corresponding conductance peaks and scattering matrix poles.

Inspection of the figures shows that the X-shaped profile appears when two scattering matrix poles cross when projected onto the real axis. (They do not cross in the complex energy plane.) The Y-shaped profile appears when $N_{\rm Y}$ jumps by two.

\section{Conclusion}
\label{conclude}

For a closed superconducting quantum dot, the distinction between topologically trivial and nontrivial is the absence or presence of a level pinned to the middle of the gap (a Majorana zero-mode). When the quantum dot is connected to a metallic reservoir, the bound states become quasi-bound, $E\mapsto E-i\gamma$, with a finite life time $\hbar/2\gamma$. The distinction between topologically trivial and nontrivial then becomes whether the number $N_{\rm Y}$ of quasi-bound states with $E=0$ is even or odd.

One can now distinguish two types of transitions \cite{Pik11}: At a topological phase transition $N_{\rm Y}$ changes by $\pm 1$ \cite{note3}. At a ``pole transition'' $N_{\rm Y}$ changes by $\pm 2$. Both types of transitions produce the same Y-shaped conductance profile of two peaks that merge and stick together for a range of parameter values --- distinct from the X-shaped profile that happens without a change in $N_{\rm Y}$.

There is a variety of methods to distinguish the pole transition from the topological phase transition \cite{Pik12}: Since $N_{\rm Y}\simeq \Gamma^{3/2}\sqrt{N}$ for $\Gamma\ll 1$, one way to suppress the pole transitions is to couple the metal to the superconductor via a small number of modes $N$ with a small transmission probability $\Gamma$. The pole transitions are a sample-specific effect, while the topological phase transition is expected to be less sensitive to microscopic details of the disorder. One would therefore not expect the pole transitions to reproduce in the same sample upon thermal cycling. If one can measure from both ends of a nanowire, one might search for correlations between the conductance peaks at the two ends. The Majorana zero-modes come in pairs, one at each end, so there should be a correlation in the conductance peaks measured at the two ends, which we would not expect to be there for the peaks due to the pole transition. 

\acknowledgments

This research was supported by the Foundation for Fundamental Research on Matter (FOM), the Netherlands Organization for Scientific Research (NWO/OCW), an ERC Synergy Grant, and the China Scholarship Council.

\appendix

\section{Factor-of-two difference in the construction of Gaussian ensembles with or without particle-hole symmetry}
\label{cpmEsym}

As we discussed in Sec.\ \ref{gaussens}, in the Gaussian ensembles of random-matrix theory the Hermitian $M\times M$ matrix $H$ has distribution
\begin{subequations}
\label{PHGaussian}
\begin{align}
&P(H)\propto\exp\left(-\frac{c}{M}\,{\rm Tr}\,H^2\right),\label{PHGaussiana}\\
&c=\frac{\beta\pi^{2}}{8\delta_0^{2}}\times\begin{cases}
2& \text{in the Wigner-Dyson ensembles},\\
1& \text{in the Altland-Zirnbauer ensembles},\\
1& \text{in the chiral ensembles}.
\end{cases}\label{PHGaussianb}
\end{align}
\end{subequations}
In each ensemble $\delta_0$ refers to the average spacing of distinct eigenvalues of $H$ in the bulk of the spectrum. For $\beta=4$ the eigenvalues have a twofold Kramers degeneracy, so there are only $M_0=M/2$ distinct eigenvalues, while for $\beta=1,2$ all $M_0=M$ eigenvalues are distinct (disregarding spin degeneracy).

We have experienced that the factor-of-two difference in the coefficient between the Wigner-Dyson (WD) and Altland-Zirnbauer (AZ) ensembles is a source of confusion. Here we hope to resolve this confusion by pointing to its origin, which is the $\pm E$ symmetry of the spectrum in the AZ ensembles (and also in the chiral ensembles, which we include for completeness). The calculation of the coefficient $c$ is a bit lengthy, with factors of two appearing at different places before the final factor remains, but we have not found a much shorter and convincing argument for the difference.

The eigenvalue distribution in the WD ensembles is \cite{Mehta,Forrester,handbook}
\begin{equation}
P(E_1,E_2,\ldots E_{M_0})\propto \prod_{1=i<j}^{M_0}|E_i-E_{j}|^{\beta}\prod_{k=1}^{M_0}e^{-\frac{c}{M_0}E_k^2},\label{PWD}
\end{equation}
where the indices $i,j,k$ range over the $M_0$ distinct eigenvalues. 

In the AZ ensembles an eigenvalue at $+E$ has a partner at $-E$, which is a distinct eigenvalue if $E\neq 0$. For the average level spacing in the bulk of the spectrum the existence of a level pinned at $E=0$ is irrelevant, so we assume that there are no such zero-modes. (This requires $M_0$ even.) The eigenvalue distribution then has the form \cite{Alt97,Iva02}
\begin{align}
&P(E_1,E_2,\ldots E_{M_0/2})\propto \prod_{1=i<j}^{M_0/2}|E_i^2-E_{j}^2|^{\beta}\nonumber\\
&\qquad\times\prod_{k=1}^{M_0/2}|E_k|^{\alpha}\exp\left(-\frac{2c}{M_0}E_k^2\right),\label{PAZ}
\end{align}
where now the indices $i,j,k$ range only over the $M_0/2$ distinct positive eigenvalues. There is a new exponent $\alpha\in\{0,1,2\}$ that governs the repulsion between eigenvalues related by $\pm E$ symmetry. This factor $|E_{k}|^\alpha$ only affects the first few levels around $E=0$, so we may ignore it for a calculation of the average level spacing in the bulk of the spectrum, effectively setting $\alpha\rightarrow 0$.

The two distributions \eqref{PWD} and \eqref{PAZ} may be written in the same form with the help of the microscopic level density
\begin{equation}
\rho(E)=\sum_{n=1}^{M_0}\delta(E-E_{n}),\label{rhoEdef}
\end{equation}
defined for each set of $M_0$ distinct energy levels. At the mean-field level, sufficient for a calculation of the density of states in the large-$M$ limit, we may assume that $\rho(E)$ is a smooth function of $E$ (Coulomb gas model \cite{Mehta}). 

The eigenvalue distribution has the form of a Gibbs distribution $P\propto\exp(-\beta U)$, with energy functional
\begin{align}
U_{\rm WD}={}&-\frac{1}{2}\int_{-\infty}^{\infty}dE\,\int_{-\infty}^{\infty}dE'\,\rho(E)\rho(E')\ln|E-E'|\nonumber\\
&+\frac{c}{\beta M_0}\int_{-\infty}^{\infty}dE\,E^{2}\rho(E),\label{UWD}
\end{align}
for the WD ensembles and
\begin{align}
U_{\rm AZ}={}&-\frac{1}{2}\int_{0}^{\infty}dE\,\int_{0}^{\infty}dE'\,\rho(E)\rho(E')\ln|E^2-E'^2|\nonumber\\
&+\frac{2c}{\beta M_0}\int_{0}^{\infty}dE\,E^{2}\rho(E)\nonumber\\
={}&-\frac{1}{4}\int_{-\infty}^{\infty}dE\,\int_{\infty}^{\infty}dE'\,\rho(E)\rho(E')\ln|E-E'|\nonumber\\
&+\frac{c}{\beta M_0}\int_{-\infty}^{\infty}dE\,E^{2}\rho(E),\label{UAZ}
\end{align}
for the AZ ensembles (at $\alpha=0$). In the second equality we used the $\pm E$ symmetry $\rho(E)=\rho(-E)$.

The mean-field density of states $\bar{\rho}(E)$ minimizes $U$ with the normalization constraint
\begin{equation}
\int_{-\infty}^{\infty}dE\,\bar{\rho}(E)=M_0.\label{barrhonorm}
\end{equation}
The normalization constraint is the same in the WD and AZ ensembles, but the minimization condition is different:
\begin{align}
&\frac{\delta U_{\rm WD}}{\delta\rho(E)}=0\Rightarrow-\int_{-\infty}^{\infty}dE'\,\bar{\rho}_{\rm WD}(E')\ln|E-E'|\nonumber\\
&\qquad\qquad+\frac{c}{\beta M_0}E^{2}={\rm constant},\label{barrhoWD}\\
&\frac{\delta U_{\rm AZ}}{\delta\rho(E)}=0\Rightarrow-\frac{1}{2}\int_{-\infty}^{\infty}dE'\,\bar{\rho}_{\rm AZ}(E')\ln|E-E'|\nonumber\\
&\qquad\qquad+\frac{c}{\beta M_0}E^{2}={\rm constant}.\label{barrhoAZ}
\end{align}
The $\pm E$ symmetry does not introduce an additional constraint on $\bar{\rho}_{\rm AZ}(E)$, since Eq.\ \eqref{barrhoAZ} automatically produces an even density.

The solution to this integral equation gives the familiar semi-circular density of states \cite{Mehta},
\begin{align}
&\bar{\rho}_{\rm WD}(E)=\frac{2c}{\pi\beta M_0}\sqrt{(\beta/c)M_0^2-E^2},\label{rhoWDsc}\\
&\bar{\rho}_{\rm AZ}(E)=\frac{4c}{\pi\beta M_0}\sqrt{(\beta/2c)M_0^2-E^2}.\label{rhoAZsc}
\end{align}
The mean level spacing near $E=0$ is $\delta_0=1/\bar{\rho}(0)$, leading to
\begin{equation}
\begin{split}
&\delta_0=\tfrac{1}{2}\pi\sqrt{\beta/c}\;\;\text{in the WD ensembles},\\
&\delta_0=\tfrac{1}{2}\pi\sqrt{\beta/2c}\;\;\text{in the AZ ensembles},
\end{split}
\end{equation}
which amounts to Eq.\ \eqref{PHGaussianb}. Notice that the additional factor-of-two arises solely from $\pm E$ symmetry of the spectrum, so it does not matter whether this is a consequence of particle-hole symmetry or of chiral symmetry.

\begin{figure}[tb]
\centerline{\includegraphics[width=1\linewidth]{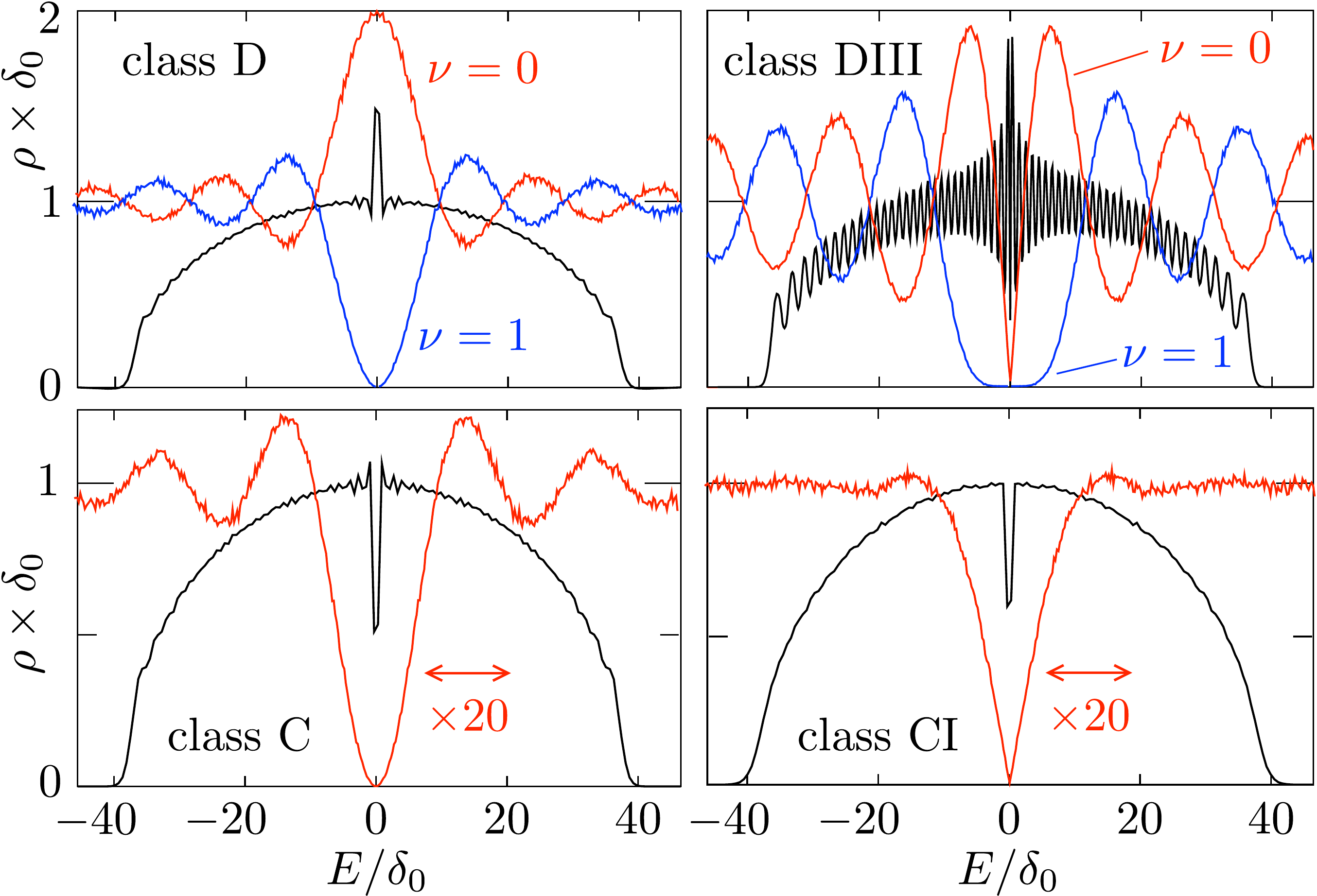}}
\caption{Black and red curves: Average density of states in the four Altland-Zirnbauer ensembles, calculated numerically for Hamiltonians of dimension $M\times M=60\times 60$ in classes C, CI, D, and $M\times M=120\times 120$ in class DIII (when each level has a twofold Kramers degeneracy; $\rho$ and $\delta_0$ refer to distinct levels). The black curve shows the full semicircle, the red curve shows the region around $E=0$ (horizontally enlarged by a factor $20$). These are all results for a topologically trivial superconductor, without a zero-mode ($\nu=0$). The blue curves (labeled $\nu=1$) show the effect of a zero-mode in class D ($M=61$) and class DIII ($M=122$). The delta-function peak from the zero-mode itself is not plotted.
}
\label{fig_DOS}
\end{figure}

To check that we have not missed a factor of two, we show in Fig.\ \ref{fig_DOS} the numerical result of an average over a large number of random Hamiltonians in each of the four Altland-Zirnbauer ensembles. The semi-circular density of states \eqref{rhoAZsc} applies away from the band center, with the expected limit $\rho\times\delta_0\rightarrow 1$ near $E=0$. 

We also see in Fig.\ \ref{fig_DOS} the anomalies at band center that we ignored in our calculation. Without a zero-mode ($\nu=0$) the density of states vanishes as $|E|^{\alpha}$ with $\alpha=2$ in class C and $\alpha=1$ in class CI and DIII \cite{Alt97}. In class D one has $\alpha=0$, which means that the $\pm E$ pairs of energy levels do not repel at the band center. The density of states then has a quadratic peak at $E=0$. The delta-function peak of a zero-mode has also an effect on the smooth part of the density of states, which for $\nu=1$ vanishes as $|E|^{\alpha+\beta}$, so as $E^2$ in class D and as $|E|^5$ in class DIII \cite{Iva02}.

\section{Altland-Zirnbauer ensembles with time-reversal symmetry}
\label{CIDIIIensembles}

For completeness and reference, we record the $\beta=1,4$ counterparts of the $\beta=2$ formulas \eqref{GaussEns} and \eqref{GaussEnsC}. These are the Altland-Zirnbauer symmetry classes CI ($\beta=1$, time-reversal with spin-rotation symmetry) and DIII ($\beta=4$, time-reversal without spin-rotation symmetry) \cite{Alt97}. The time-reversal symmetry conditions on the Hamiltonian matrix are
\begin{equation}
\begin{split}
&H=H^{\ast}\;\;{\rm for}\;\;\beta=1,\\
&H=\sigma_{y}H^{\ast}\sigma_{y}\;\;{\rm for}\;\;\beta=4.
\end{split}\label{trsconditions}
\end{equation}
The Pauli matrix $\sigma_y$ acts on the spin degree of freedom --- the Pauli matrices $\tau_{i}$ we used earlier acted on the electron-hole degree of freedom.

A compact representation can be given if we use the electron-hole basis for $\beta=1$ and the Majorana basis for $\beta=4$. The matrix elements of the Hamiltonian can then be represented by Pauli matrices:
\begin{equation}
\begin{split}
&H_{nm}=a_{nm}\tau_{x}+b_{nm}\tau_{z}\;\;{\rm for}\;\;\beta=1,\\
&H_{nm}=ic_{nm}\sigma_{x}+id_{nm}\sigma_{z}\;\;{\rm for}\;\;\beta=4,
\end{split}\label{Hnmbeta14}
\end{equation}
with real coefficients $a,b,c,d$. Notice that $iH$ for $\beta=1$ is quaternion, so this class CI ensemble is a subset of the class C ensemble. Similarly, because $iH$ is real for $\beta=4$, this class DIII ensemble is a subset of class D.

Hermiticity of $H$ requires that the off-diagonal elements are related by $a_{nm}=a_{mn}$, $b_{nm}=b_{mn}$, $c_{nm}=-c_{mn}$, $d_{nm}=-d_{mn}$. On the diagonal $c_{nn}=d_{nn}=0$. The indices $n,m$ range from $1$ to $M/2$, for an $M\times M$ matrix $H$. (The dimensionality is necessarily even to accomodate the Pauli matrices.) For $\beta=4$ there is a twofold Kramers degeneracy of the energy levels, so only $M/2$ eigenvalues of $H$ are distinct. For $\beta=1$ all $M$ eigenvalues are distinct (the spin degeneracy that exists in class C, CI is not included in $M$). The mean level spacing $\delta_0$ refers to the distinct eigenvalues.

Combination of Eq.\ \eqref{Hnmbeta14} with Eqs.\ \eqref{Eq02} and \eqref{cdef} gives the probability distribution of the independent matrix elements in the Gaussian ensemble:
\begin{align}
&P(\{H_{nm}\})\propto\prod_{n=1}^{M/2}\exp\left(-\frac{\pi^{2}}{4M\delta_0^{2}}(a_{nn}^2+b_{nn}^2)\right) \nonumber\\
&\quad\times  \prod_{1=n<m}^{M/2}\exp\left(-\frac{\pi^{2}}{2M\delta_0^{2}}(a_{nm}^2+b_{nm}^2)\right),\label{GaussEnsCI}
\end{align}
for $\beta=1$, class CI, and
\begin{align}
&P(\{H_{nm}\})\propto \prod_{1=n<m}^{M/2}\exp\left(-\frac{2\pi^{2}}{M\delta_0^{2}}(c_{nm}^2+d_{nm}^2)\right),\label{GaussEnsDIII}
\end{align}
for $\beta=4$, class DIII.

\section{Mapping of the pole statistics problem onto the eigenvalue statistics problem of truncated orthogonal matrices}
\label{orthogonalmapping}

We show how the result \eqref{rho0result} for the density profile of imaginary poles of the scattering matrix follows from the known distribution of real eigenvalues of truncated orthogonal matrices \cite{Kho10} --- for the case $\Gamma=1$ of ballistic coupling.

Following Ref.\ \onlinecite{Bro99} we construct the $N \times N$ energy-dependent unitary scattering matrix $S(E)$ in terms of an $M\times M$ energy-independent orthogonal matrix $O$,
\begin{equation}
  S(E) = {\cal P} O (e^{-2 \pi i E/M \delta_0} +  {\cal R} O)^{-1}{\cal P}^{\rm T}. \label{eq:SU1}
\end{equation}
The rectangular $N\times M$ matrix ${\cal P}$ has elements ${\cal P}_{nm} =\delta_{nm}$ and ${\cal R} = 1 - {\cal P}^{\rm T}{\cal P}$. The $M\times M$ Hermitian matrix $H$ is related to $O$ via a Cayley transform,
\begin{equation}
O = \frac{\pi H/M\delta_0+i}{\pi H/M\delta_0-i}
\Leftrightarrow H=\frac{i M \delta_0}{\pi}\,\frac{O+1}{O-1}.
\label{eq:H0def}
\end{equation}
Eq.\ \eqref{eq:H0def} with $O$ uniformly distributed according to the Haar measure in ${\rm SO}(N)$ produces the Gaussian distribution \eqref{Eq02} for $H$, in the low-energy range $|E|\lesssim N\delta_0\ll M\delta_0$. Furthermore, in this low-energy range the scattering matrix \eqref{eq:SU1} is related to $H$ by Eq.\ \eqref{Eq03} with ballistic coupling matrix $W = {\cal P}^{\rm T} (M\delta_0 /\pi^2)^{1/2}$.

A pole $\varepsilon=-i\gamma$ of $S(\varepsilon)$ on the imaginary axis corresponds to a real eigenvalue
\begin{equation}
x=e^{-2\pi\gamma/M\delta_0}\label{xgammarelation}
\end{equation}
of the $(M-N)\times(M-N)$ matrix $\tilde{O}={\cal R}O{\cal R}$ obtained from the orthogonal matrix $O$ by deleting the first $N$ rows and columns. For $M\gg 1$ the $x$-dependent density $\tilde{\rho}_0(x)$ is given by \cite{Kho10}
\begin{equation}
\tilde{\rho}_0(x)=\frac{1}{B(N/2,1/2)}\frac{1}{1-x^2},\;\;x^2<1-N/M,\label{rho0x}
\end{equation}
with $B(a,b)$ the beta function. 

Using Eq.\ \eqref{xgammarelation} we thus arrive for $N\ll M$ at the $\gamma$-dependent density
\begin{equation}
\rho_0(\gamma)=\frac{1}{B(N/2,1/2)}\frac{1}{2\gamma},\;\;\gamma>N\delta_0/4\pi.\label{rho0gammaresult}
\end{equation}
Eq.\ \eqref{rho0result} with $\Gamma=1$ results if we also assume that $N\gg 1$, so that we may approximate $B(N/2,1/2)\approx (2\pi/N)^{1/2}$.


\begin{thebibliography}{99}
\bibitem{note3} Since Majorana zero-modes always appear in pairs, the change from $M$ even to $M$ odd necessarily involves some external system that can absorb one of the two Majoranas. For example, this could be a nanowire coupled at one end to the quantum dot, so that the Majorana zero-mode inside the quantum dot is paired with the zero-mode at the other end of the wire. At the transition from $M$ odd to even the two Majoranas merge because the gap in the nanowire closes.
\bibitem{note2} The usual Kramers degeneracy refers to the eigenvalues of a Hermitian matrix that commutes with an anti-unitary operator squaring to $-1$. Here the matrix is not Hermitian, but the degeneracy still applies to real eigenvalues.
\bibitem{note4} The logarithmic divergence in Eq.\ \eqref{squarerootlaw} for $\Gamma=1$ is cut-off by the finite dimension $M$ of the Hamiltonian, so that $\langle N_Y\rangle\simeq N^{1/2}\ln(M/N)$ for ballistic coupling. This spurious $M$-dependence does not exist for $\Gamma<1$.
\end{thebibliography}
\end{document}